\documentclass[aps,prd,preprint,superscriptaddress,showpacs,showkeys]{revtex4}
\usepackage{amssymb,amsmath}
\usepackage{graphicx} 
\usepackage{dcolumn}  
\usepackage{epsfig}   
\usepackage{pstricks}
\usepackage{ifpdf}
\newcommand{\Tr}{ {\mathrm{Tr}\, }}

\begin{document}


\title{On the non-perturbative realization of QCD gauge-invariance}


%
%

\author{H.M. Fried}
\affiliation{Physics Department, Brown University, Providence, RI 02912, USA}
\email[]{Herb Fried fried@het.brown.edu}
\author{T. Grandou}
\affiliation{Universit\'{e} de Nice-Sophia Antipolis,\\ Institut Non Lin\'{e}aire de Nice, UMR CNRS 7335; 1361 routes des Lucioles, 06560 Valbonne, France}
\email[]{Thierry.Grandou@inln.cnrs.fr}
\author{R. Hofmann}
\affiliation{Institut f\"ur Theoretische Physik\\ 
Universit\"at Heidelberg\\ 
Philosophenweg 16\\ 
69120 HEIDELBERG}
\email[]{r.hofmann@thphys.uni-heidelberg.de}


\date{\today}

\begin{abstract} A few years ago the use of standard functional manipulations was demonstrated to imply an unexpected property satisfied by the fermionic Green's functions of QCD: \textit{effective locality}. This feature of QCD is non-perturbative as it results from a full integration of the gluonic degrees of freedom. In this paper, previous derivations of effective locality are reviewed, corrected, and enhanced. Focussing on the way non-abelian gauge invariance is realized in the non-perturbative regime of QCD, the deeper meaning of effective locality is discussed. \end{abstract}


\maketitle

\section{\label{SEC:1}Introduction}

In some recent articles~\cite{QCD1,QCD-II,QCD5, QCD6, QCD5'} a property of the non-perturbative fermionic Green's functions in QCD was put forward under the name {\textit{effective locality}}. This property
can be summarized as follows. 
\par
\textit{For any fermionic $2n$-point Green's functions and related amplitudes, the
full gauge-fixed sum of cubic and quartic gluonic interactions,
fermionic loops included, results in a local contact-type interaction. This local interaction is
mediated by a tensorial field which is antisymmetric both in Lorentz and
color indices. Moreover, the resulting sum appears to be fully gauge-fixing independent, that is, gauge-invariant.}
\par
 This is a non-expected result because integrations of elementary degrees of freedom ordinarily result in highly non-local 
 structures. The
`effective locality' denomination accounts for this unusual circumstance. Notice that in the pure euclidean Yang Mills case, and up to the first non-trivial orders of a semi-classical expansion, effective locality was observed a welcome property in an attempt to construct a formulation dual to the original Yang Mills theory \cite{RefF}.
\par
Now, apart from a \textit{supersymmetric} extension, QCD is not known to possess any dual formulation, and the full effective locality functional expressions certainly attests to this difficulty. However, like in the pure Yang Mills situation of \cite{RefF}, effective locality is a useful means to learn about non-perturbative physics in QCD, and this based on first principles. 
\par\medskip\medskip
This paper is organized as follows. In the next Section a series of standard functional manipulations is used to construct the generating functionals of QED and QCD, and to contrast them at one essential point. In Section \ref{SEC:3}, the functional statement of effective locality is given.  At first, it is illustrated for the simpler case of \textit{eikonal} and \textit{quenching} approximations  to QCD. Then it is shown that effective locality holds true also in the full non-approximated QCD case. Furthermore, it is shown that this property is even independent of the (Fradkin's) representations used for the fermionic propagator in a background $A_\mu$- field configuration. Section \ref{SEC:4} summarizes our results and discusses the meaning of effective locality further. 

 \section{\label{SEC:2}Generating functionals}

 \par
 In QED it is known for long  that manifest covariance and manifest gauge invariance are competing aspects of a generating functional construction. In the present Section it is shown how the conventional Schwinger solution for the generating functionals of QED and QCD, lead to expressions which, beyond appearances, differ on an essential point. Beginning with QED, the free photonic Lagrangian density reads,
\begin{equation}\label{Eq:1}
\mathcal{L}_{0} = - \frac{1}{4} {f}^{\mu \nu} {f}_{\mu \nu} =  - \frac{1}{4}\left( \partial^{\mu} A^{\nu} -  \partial^{\nu} A^{\mu} \right) \left( \partial_{\mu} A_{\nu} -  \partial_{\nu} A_{\mu} \right).
\end{equation}
The corresponding action may be written as
\begin{eqnarray}\label{Eq:2}
\int{\mathrm{d}^{4}x \, \mathcal{L}_{0}}  &=& - \frac{1}{2} \int{ A^{\mu} \, \left( -\partial^{2}\right) \, A_{\mu} } + \frac{1}{2} \int{ \left( \partial^{\mu} A_{\mu} \right)^{2} },
\end{eqnarray}

\noindent The difficulty of maintaining both manifest gauge invariance and manifest Lorentz covariance can be discussed at this stage.  What has typically been done since the original days of Fermi, who simply neglected the inconvenient term $\left( \partial_{\mu} A^{\mu} \right)^{2}$, is to use it in order to define a relativistic gauge in which all calculations exhibit manifest Lorentz covariance, while relying upon strict charge conservation to maintain an effective gauge invariance of the theory. The choice of a relativistic gauge can be arranged in various ways. The simplest functional way is to multiply the inconvenient term by a real parameter $\lambda$, and treat it as an interaction term \cite{schwinger}. For example, starting from the free-field, ($\lambda = 0$, Feynman) propagator ${D}_{F}{}^{(0)}_{\mu \nu} =g_{\mu \nu} \, {D}_{F}$, where $(-\partial^{2}) \, {D}_{F} = \delta^{4}$, one has the free-field generating functional,
\begin{equation}\label{Eq:3}
{Z}_{0}^{(0)}\{j\} = \exp{\left\{ \frac{i}{2} \int{j \cdot
{D}_{\mathrm{F}}^{(0)} \cdot j} \right\} }
\end{equation}
\noindent Then, operating upon it with the 'interaction' $\lambda$-term, a new free-field generating functional is obtained.
\begin{eqnarray}\label{Eq:4}
{Z}_{0}^{(\zeta)}\{j\} &=& \left. e^{\frac{i}{2} \lambda \int{ \left( \partial^{\mu} A_{\mu} \right)^{2} }} \right|_{A \rightarrow \frac{1}{i}\frac{\delta}{\delta j}} \cdot e^{\frac{i}{2} \int{j \cdot {D}_{\mathrm{F}}^{(0)} \cdot j} }\nonumber \\  &=& e^{-\frac{i}{2} \Tr\ln{\left[ 1 - \lambda\,( {\partial\otimes \partial/\partial^{2}}) \right]}}\ e^{ \frac{i}{2} \int{j \cdot
{D}_{\mathrm{F}}^{(\zeta)} \cdot j} } \,,
\end{eqnarray}
\noindent where	
\begin{equation}\label{DF}{D}_{\mathrm{F}\mu \nu}^{(\zeta)} = \left[ g_{\mu \nu} - \zeta \partial_{\mu} \partial_{\nu} / \partial^{2} \right] \, {D}_{\mathrm{F}}\,, \ \ \ \ \  \zeta = \frac{\lambda}{1 - \lambda}\,.\end{equation}  Note that the functional differential operation of (\ref{Eq:4}) is fully equivalent to a bosonic gaussian functional integration with the advantage that it does not require the specification of any (infinite) normalization constant. In (\ref{Eq:4}), the Tr-Log term is an infinite phase factor representing the sum of the vacuum energies generated by longitudinal and time-like photons, with a weight $\lambda$ arbitrarily inserted. This quantity can be removed by an appropriate version of normal ordering, or more simply, be absorbed into an overall normalization constant.
\par\medskip Including now the fermionic interaction, $\mathcal{L}_{\mathrm{int}} = -ig \bar{\psi} \gamma \cdot A \psi$, and the `gauge interaction' $\frac{1}{2} \lambda \left( \partial_{\mu} A_{\mu} \right)^{2}$, the standard Schwinger solution for the generating functional is obtained as,
\begin{eqnarray}\label{Eq:5}
{Z}_{\mathrm{QED}}^{(\zeta)}[j,\eta,\bar{\eta}] = \mathcal{N} \left. e^{i\int{\bar{\eta} \cdot {G}_{\mathrm{F}}[A] \cdot \eta} + {L}[A] + \frac{i}{2} \lambda \int{ \left( \partial^{\mu} A_{\mu} \right)^{2} }} \right|_{A \rightarrow \frac{1}{i}\frac{\delta}{\delta j}} \cdot e^{\frac{i}{2} \int{j \cdot {D}_{\mathrm{F}}^{(0)} \cdot j} }
\end{eqnarray}

\noindent with, \begin{equation}{G}_{F}[A] = [\gamma \cdot (\partial - i e A)-m]^{-1}\,,\ \ \ \ {L}[A] = \Tr\ln{\left[ 1 - ig \gamma \cdot A {S}_{\mathrm{F}} \right]}\,,\ \ \ \ {S}_{\mathrm{F}} = {G}_{\mathrm{F}}[0]\,,\end{equation} and where the phase factor of (\ref{Eq:4}) has been absorbed into $\mathcal{N}$.  A convenient re-arrangement of (\ref{Eq:5}) uses the identity, satisfied by any polynomial and exponential functional $\mathcal{F}[A]$ \cite{RefL},
\begin{eqnarray}\label{Eq:6}
\mathcal{F}\left[ \frac{1}{i} \frac{\delta}{\delta j} \right] \cdot e^{\frac{i}{2} \int{j \cdot {D}_{\mathrm{F}}^{(\zeta)} \cdot j} } = e^{\frac{i}{2} \int{j \cdot {D}_{\mathrm{F}}^{(\zeta)} \cdot j} } \cdot \left. e^{\mathfrak{D}_{A}} \cdot \mathcal{F}[A] \right|_{A = \int{{D}_{\mathrm{F}}^{(\zeta)} \cdot j} }
\end{eqnarray}

\noindent where $\mathfrak{D}^{(\zeta)} _{A}$, is the so-called \textit{linkage operator}, 
$\mathfrak{D}^{(\zeta)} _{A} =  - \frac{i}{2} \int\mathrm{d}^4x\int\mathrm{d}^4y{\frac{\delta}{\delta A(x)} \cdot  {D}_{\mathrm{F}}^{(\zeta)}(x-y) \cdot \frac{\delta}{\delta A(y)} }$.
This identity can be used to eventually write,
\begin{eqnarray}\label{Eq:7}
{Z}_{\mathrm{QED}}^{(\zeta)}[j,\eta,\bar{\eta}] = \mathcal{N} \, e^{\frac{i}{2} \int{j \cdot {D}_{\mathrm{F}}^{(\zeta)} \cdot j} } \cdot \left. e^{\mathfrak{D}_{A}^{(\zeta)}} \cdot e^{i\int{\bar{\eta} \cdot {G}_{\mathrm{F}}[A] \cdot \eta} + {L}[A]} \right|_{A = \int{{D}_{\mathrm{F}}^{(\zeta)} \cdot j}}\,.
\end{eqnarray}
\noindent This is nothing but the formal solution for the QED generating functional which would result from a linear and covariant gauge fixing term $\frac{-1}{2(1-\lambda)}(\partial\cdot A)^2$. Accordingly, the Green's functions given by (\ref{Eq:7}) cannot display gauge invariance: The bare photon propagator is explicitly gauge-dependent and it is only through the control of Ward identities that 
radiative corrections invoking this propagator are proven to be gauge invariant. Likewise, using such a gauge dependent photon propagator does not prevent properly defined $S$- matrix elements from being gauge-invariant, by virtue of the LSZ reduction formula and \emph{equivalence theorem} \cite{{RefO'},{IZ}}. The point to be retained here is that recovering gauge invariance goes along some circuitous path. \par\bigskip\medskip
For QCD, one starts from a Lagrangian density,
\begin{equation}\label{Eq:8}
\mathcal{L}_{\mathrm{QCD}} = - \frac{1}{4} {F}_{\mu \nu}^{a} {{F}^{\mu \nu}}^{a} - \bar{\psi} \cdot [ \gamma^{\mu} \, (\partial_{\mu} - i g A_{\mu}^{a} \lambda^{a})-m] \cdot \psi,
\end{equation}

\noindent where, in ${F}_{\mu \nu}^{a} = \partial_{\mu} A_{\nu}^{a} -  \partial_{\nu} A_{\mu}^{a} + g f^{abc} A_{\mu}^{b} A_{\nu}^{c}$, the numbers $f^{abc}$ are the customary structure constants of the $SU_c(3)$ Lie algebra. 
 
 One can begin by identifying the free and interacting gluonic pieces,
\begin{equation}\label{Eq:9}
-\frac{1}{4} \int{{F}^{2}} = -\frac{1}{4} \int{{f}^{2}} - \frac{1}{4} \int{\left[{F}^{2} - {f}^{2} \right]} \equiv -\frac{1}{4} \int{{f}^{2}} + \int{\mathcal{L}'[A]},
\end{equation}
\noindent with,
\begin{equation}\label{glue}
{f}_{\mu \nu}^{a} \equiv \partial_{\mu} A_{\nu}^{a} -  \partial_{\nu} A_{\mu}^{a}\\,,\ \ \ \ \mathcal{L}'[A] =  -\frac{1}{4} \left(2 {f}_{\mu \nu}^{a} +  g f^{abc} A_{\mu}^{b} A_{\nu}^{c} \right) \, \left(g f^{abc} A^{\mu}_{b} A^{\nu}_{c} \right)\,.\end{equation} In analogy with QED, after an integration-by-parts, one can write,
\begin{equation}\label{Eq:10}
-\frac{1}{4} \int{{F}^{2}} = - \frac{1}{2} \int{ A^{\mu}_{a} \, \left( -\partial^{2}\right) \, A_{\mu}^{a} } + \frac{1}{2} \int{ \left( \partial^{\mu} A_{\mu}^{a} \right)^{2} } + \int{\mathcal{L}'[A]},
\end{equation}
\noindent and focus on the gluonic part of QCD only.
	In order to select a particular covariant gauge propagator, one can multiply the 2nd right hand side term of (\ref{Eq:10}) by $\lambda$, and include this term as part of the interaction, obtaining the generating functional,
\begin{eqnarray}\label{Eq:11}
{Z}_{\mathrm{}}^{(\zeta)}[j] = \mathcal{N} \, e^{i \int{\mathcal{L}'\left[\frac{1}{i} \, \frac{\delta}{\delta j} \right]}} \  e^{\frac{i}{2} \lambda \int{\frac{\delta}{\delta j_{\mu}} \, \partial_{\mu} \partial_{\nu} \, \frac{\delta}{\delta j_{\nu}}} } \  e^{\frac{i}{2} \int{j \  {D}_{\mathrm{F}}^{(0)} \cdot j} },
\end{eqnarray}
\noindent that is,
\begin{eqnarray}\label{Eq:12}
{Z}_{\mathrm{}}^{(\zeta)}[j] = \mathcal{N} \, e^{i \int{\mathcal{L}'\left[\frac{1}{i} \, \frac{\delta}{\delta j} \right]}} \  e^{\frac{i}{2} \int{j \cdot {D}_{\mathrm{F}}^{(\zeta)} \cdot j} }
\end{eqnarray}

\noindent with the determinantal phase factor of (\ref{Eq:4}) included in the normalization $\mathcal{N}$, and, as the free gluon propagator, the expression ${D}^{(\zeta)}_{\mathrm{F}}$ of (\ref{DF}) multiplied by the color factor of $\delta^{ab}$.
Fermionic variables can be re-inserted now to give,
\begin{equation}\label{ZQCD}
Z^{(\zeta)}_{QCD}[j,\bar{\eta}, \eta]=\mathcal{N} \, e^{i \int{\mathcal{L'_{QCD}}\left[\frac{1}{i} \, \frac{\delta}{\delta j} \right]}} \  e^{\frac{i}{2} \int{j \cdot {D}_{\mathrm{F}}^{(\zeta)} \cdot j} }\end{equation}with,
\begin{equation}\mathcal{L'}_{\mathrm{QCD}}[A]=\mathcal{L'}[A]+\bar{\eta}\,G_{\mathrm{F}}[A]\,\eta +L[A]\,.
\end{equation}
In view of (\ref{DF}), however, it is to be noted that all choices of $\lambda$ are possible at the exception of $\lambda = 1$, for that choice leads to an ill-defined gluon propagator ($\zeta \rightarrow \infty$).  This is an unfortunate situation, because, as is clear from (\ref{Eq:10}), the choice $\lambda = 1$ is precisely the one corresponding to manifest gauge invariance in QCD, as well as in QED.
\par
At this point, though, the situation in QCD differs from that of QED in a meaningful way. In effect, using for $\int{\mathcal{L}'[A]}$ the expression given by (\ref{Eq:10}),
and omitting the quark variables momentarily, one can write,
\begin{eqnarray}\label{Eq:14}
{Z}_{\mathrm{}}^{(\zeta)}[j] = \mathcal{N} \, \left. e^{-\frac{i}{4} \int{{F}^{2}} - \frac{i}{2} (1 - \lambda) \int{ \left( \partial^{\mu} A_{\mu}^{a} \right)^{2}} + \frac{i}{2} \int{ A_{\mu}^{a} \, \left( -\partial^{2}\right) \, A^{\mu}_{a} } } \right|_{A \rightarrow \frac{1}{i} \, \frac{\delta}{\delta j} } \, e^{\frac{i}{2} \int{j \cdot {D}_{\mathrm{F}}^{(0)} \cdot j} }\,,
\end{eqnarray}
which renders obvious that now, contrarily to the QED case, the choice $\lambda =1$ can be made. For the full QCD generating functional, this gives,
\begin{eqnarray}\label{Eq:16}
{Z}_{\mathrm{QCD}}[j, \bar{\eta}, \eta] &=& \mathcal{N}e^{\frac{i}{2} \int{j \cdot {D}_{\mathrm{F}}^{(0)} \cdot j}}\nonumber\\ &\times& \left. e^{- \frac{i}{2} \int{\frac{\delta}{\delta A} \cdot {D}_{\mathrm{F}}^{(0)} \cdot \frac{\delta}{\delta A} } } \cdot e^{-\frac{i}{4} \int{{F}^{2}} + \frac{i}{2} \int{ A \cdot \left( -\partial^{2}\right) \cdot A} } \cdot e^{i\int{\bar{\eta} \cdot {G}_{\mathrm{F}}[A] \cdot \eta} + {L}[A]}\right|_{A = \int{{D}_{\mathrm{F}}^{(0)} \cdot j} }
\end{eqnarray}where the identity (\ref{Eq:6}) has been used. The choice of $\lambda=1$, complying both with manifest Lorentz covariance and gauge invariance, seems to have selected the Feynman field function $ \left. {D}_{\mathrm{F}}^{(0)}\right|^{ab}_{\mu\nu}$. 
\par
However, just by adding and subtracting a covariant gauge-fixing term of $\frac{1}{2\zeta}(\partial\cdot {A})^2$ in the right hand side of (\ref{Eq:10}), an alternate expression results for the generating functional (\ref{Eq:16}), which reads now,
\begin{eqnarray}\label{Eq:16'}
{Z}_{\mathrm{QCD}}[j, \bar{\eta}, \eta] &=& \mathcal{N}e^{\frac{i}{2} \int{j \cdot {D}_{\mathrm{F}}^{(\zeta)} \cdot j}}\nonumber\\ &\times& \left. e^{- \frac{i}{2} \int{\frac{\delta}{\delta A} \cdot {D}_{\mathrm{F}}^{(\zeta)} \cdot \frac{\delta}{\delta A} } } \cdot e^{-\frac{i}{4} \int{{F}^{2}} + \frac{i}{2} \int{ A \cdot \left({ {D}_{\mathrm{F}}^{(\zeta)}}\right)^{-1} \cdot A} } \cdot e^{i\int{\bar{\eta} \cdot {G}_{\mathrm{F}}[A] \cdot \eta} + {L}[A]}\right|_{A = \int{{D}_{\mathrm{F}}^{(\zeta)}\cdot j} }\,,
\end{eqnarray}where the (inverse) covariant propagator,
\begin{equation}\label{covprop}
\left({{D}_{\mathrm{F}}^{(\zeta)}}^{-1}\right)_{\mu \nu}^{a b} = -i \delta^{a b} \, \left[ g_{\mu \nu} \, \partial^{2} + \left(\frac{1}{\zeta} -1 \right) \partial_{\mu} \partial_{\nu} \right]\,,
\end{equation}generalizes the previous Feynman propagator, $ \left. {D}_{\mathrm{F}}^{(0)}\right|^{ab}_{\mu\nu}$. The same manipulation can be done also with a non-covariant gauge-fixing term of $\frac{1}{2\zeta}(\mathbf{n}\cdot {A})^2$, where $\mathbf{n}$ is a lightlike vector, $\mathbf{n}^2=0$. Then, in the limit of $\zeta\rightarrow 0$, one gets the \emph{axial planar} gauge field function, propagating the 2 gauge field physical degrees of freedom only,
\begin{equation}\label{axialprop}
\left({{D}_{\mathrm{F}}^{(\mathbf{n})}}^{}\right)_{\mu \nu}^{a b}(K) = - i\delta^{a b} \, \left(\frac{g_{\mu\nu}}{K^2+i\varepsilon}-\frac{K_\mu n_\nu+K_\nu n_\mu}{(K^2+i\varepsilon)(K\cdot \mathbf{n})}\right)\,,\ \Tr[g_{\mu\nu}-\frac{K_\mu n_\nu+K_\nu n_\mu}{K\cdot \mathbf{n}}]=2\,.
\end{equation}The associated generating functional reads then,
\begin{eqnarray}\label{Eq:16''}
{Z}_{\mathrm{QCD}}[j, \bar{\eta}, \eta] &=& \mathcal{N}e^{\frac{i}{2} \int{j \cdot {D}_{\mathrm{F}}^{(\mathbf{n})} \cdot j}}\nonumber\\ &\times& \left. e^{- \frac{i}{2} \int{\frac{\delta}{\delta A} \cdot {D}_{\mathrm{F}}^{(\mathbf{n})} \cdot \frac{\delta}{\delta A} } } \cdot e^{-\frac{i}{4} \int{{F}^{2}} + \frac{i}{2} \int{ A \cdot \left({ {D}_{\mathrm{F}}^{(\mathbf{n})}}\right)^{-1} \cdot A} } \cdot e^{i\int{\bar{\eta} \cdot {G}_{\mathrm{F}}[A] \cdot \eta} + {L}[A]}\right|_{A = \int{{D}_{\mathrm{F}}^{(\mathbf{n})}\cdot j} }.
\end{eqnarray} \par
Clearly, as many identical forms as desired can be obtained for ${Z}_{\mathrm{QCD}}[j, \bar{\eta}, \eta]$, differing only the bare gluonic field function in use, $D_F^{(n)}$, $D_F^{(\zeta)}$, etc..
\vfill\eject
\par
Some comments are in order. 
\par
(i) In the case of QED, as advertised above, the generating functional (\ref{Eq:7}) is the standard one, in a $\zeta$-dependent covariant gauge. It is easy to check that an expansion in the coupling constant reproduces both the conventional Feynman graphs and rules of Perturbation Theory in that gauge. Subsequently, gauge-invariance is controlled along the somewhat \emph{circuitous} steps which are reviewed after Equation (\ref{Eq:7}).
\par
(ii) \emph{Mutatis mutandis}, the same applies to QCD which possesses an asymptotically free phase where Perurbation Theory can be used. In covariant gauges, though, it is well known that extra complications come about. Auxiliary Faddeev-Popov fields must be introduced so that both unitarity and gauge invariance be recovered through the BRS(T) symmetry of the full Lagrangian. Slavnov-Taylor identities derive from this symmetry in the same (though much more involved) way as Ward identites derive from the U(1) symmetry of QED, and charge conservation is guaranteed in either cases. So long as gauge fields and/or the coupling are not too strong, the overall covariant construction is well defined as it avoids the unsolved Gribov's copies issue \cite{Thess}. The two points which deserve to be emphasized at this level are, first, that the so constructed generating functionals do not really extend beyond the realm of Perturbation Theory \cite{Becchi}, and in the second place, that the control of gauge-invariance goes along the same circuitous path as in QED.
\par
(iii) For QCD, however, the same formal manipulations as used for QED, and leading to the generating functional form of (\ref{Eq:16}),  may obscure the crucial difference with QED. This is because contrarily to (\ref{Eq:7}) in the case of QED, (\ref{Eq:16}) is \emph{not} the Green's function generating functional of QCD in the Feynman gauge. And likewise, (\ref{Eq:16'}) and (\ref{Eq:16''}) are not the Green's function generating functionals of QCD in the $\zeta$-covariant and axial planar gauges either. The difference is easily disclosed by the trick of adding/subtracting a certain gauge-fixing density term such as those leading to (\ref{Eq:16'}) or (\ref{Eq:16''}), as (\ref{Eq:16}) is but a special case of (\ref{Eq:16'}).\par
Now, adding and subtracting a gauge-fixing term preserves the full gauge-invariance of the original Lagrangian density. Whereas the functional operations leading to (\ref{Eq:7}) effectively break the gauge invariance of the QED Lagrangian density (the choice of $\lambda=1$ cannot be performed), and in this way just amount to a construction of the generating functional in a given linear covariant gauge, the same functional operations do not produce the same output over the non abelian structure of QCD which maintains its full original gauge invariance (the choice of $\lambda=1$ can be performed).\par\medskip
That is, in spite of the propagators appearing in (\ref{Eq:16}), (\ref{Eq:16'}) and (\ref{Eq:16''}), and which in Perturbation Theory do \emph{always} refer to covariant and axial planar gauge- fixing terms, that is to explicit gauge symmetry breaking, no gauge selection is actually performed in the generating functional constructions of (\ref{Eq:16}), (\ref{Eq:16'}) and (\ref{Eq:16''}). As demonstrated below, this surprising \emph{hiatus} finds its resolution in the property of effective locality, that is in the fact that, at the exception of the trivial and irrelevant first factors, \textit{the fermionic momenta} of (\ref{Eq:16}), (\ref{Eq:16'}) and (\ref{Eq:16''}) retain no dependence on the bare gauge field propagator one starts from.
 \section{\label{SEC:3}Effective locality}
  This property surfaces at the level of fermionic Green's functions once functional differentiations of (\ref{Eq:16}) with respect to the sources $\bar{\eta}$, $\eta$ are taken which subsequently are cancelled out, $\bar{\eta}=\eta=j_\mu^a=0$. In order to derive the property of effective locality, it is appropriate to introduce a `linearization' of the $\int F^2$ term which appears in the right hand sides of all three expressions (\ref{Eq:16}), (\ref{Eq:16'}) and (\ref{Eq:16''}). This can be achieved by using the representation, \cite{RefF,Halpern1977a,Halpern1977b},
\begin{equation}\label{Eq:17}
e^{-\frac{i}{4} \int{{F}^{2}}} = \mathcal{N}' \, \int{\mathrm{d}[\chi] \, e^{ \frac{i}{4} \int{ \left(\chi_{\mu \nu}^{a}\right)^{2} + \frac{i}{2} \int{ \chi^{\mu \nu}_{a} {F}_{\mu \nu}^{a}} } } }
\end{equation}
\noindent where, 
\begin{equation}\int{\mathrm{d}[\chi]} = \prod_{z} \prod_{a} \prod_{\mu <\nu} \int{\mathrm{d}[\chi_{\mu \nu}^{a}](z)}\,.
\end{equation} As usual, space-time is broken up into small cells of infinitesimal size $\delta^{4}$ about each point $z$, and $\mathcal{N}'$ is a normalization constant so chosen that the right hand side of (\ref{Eq:17}) becomes equal to unity as ${F}_{\mu \nu}^{a} \rightarrow 0$.  In this way, the generating functional (\ref{Eq:16}) may be re-written as (with $\mathcal{N}' \cdot \mathcal{N} \equiv \mathcal{N}''\rightarrow \mathcal{N}$),
\begin{eqnarray}\label{Eq:18}
{Z}_{\mathrm{QCD}}[j,\bar{\eta},\eta] = \mathcal{N} \, \int{\mathrm{d}[\chi] \, e^{ \frac{i}{4} \int{ \chi^{2} }} } \, \left. e^{\mathfrak{D}_{A}^{(0)}} \cdot e^{-\frac{i}{2} \int{\chi \cdot {F} + \frac{i}{2} \int{ A \cdot \left( -\partial^{2}\right) \cdot A} }} \cdot e^{i\int{\bar{\eta} \cdot {G}_{\mathrm{F}}[A] \cdot \eta} + {L}[A]}\right|_{A = \int{{D}_{\mathrm{F}}^{(0)} \cdot j} }
\end{eqnarray}
\noindent where,  
\begin{equation}\label{link}\mathfrak{D}_{A}^{(0)} = - \frac{i}{2} \int \mathrm{d}^4x\int \mathrm{d}^4y\ {\frac{\delta}{\delta A^a_\mu(x)} \left. {D}_{\mathrm{F}}^{(0)}\right|^{ab}_{\mu\nu}(x-y)\, \frac{\delta}{\delta A_\nu^b(y)}}\,,\end{equation} and where the integration on $\chi$ was permuted with the action of the functional differentiations prescribed by $\exp\,\mathfrak{D}_{A}^{(0)}$.
\par
As can be noticed, using the representation (\ref{Eq:17}) breaks the manifest gauge invariance of the left hand side. This can be remedied in several ways. For example, one can gauge the $\chi^a_{\mu\nu}$-fields so that the term $(\chi\cdot F)$ recovers a manifest invariance. In \cite{RefF}, this route was followed successfully in order to reach a (semi-classical) formulation dual to the pure euclidean Yang Mills theory. Another possibility is to complete the $\chi^a_{\mu\nu}$ functional integrations in an exact way, so as to be guaranteed to deal with the full invariant left hand side of (\ref{Eq:17}). At eikonal and quenching approximations at least, this goal has been achieved in the strong coupling limit, by using \emph{Random Matrix} calculus \cite{QCD6,RefI}. 
\par
However, so long as the issue of gauge invariance is concerned only, there is no need to cope with any of the two above possibilities and this remarkable simplification is due to the property of effective locality displayed by the \emph{fermionic momenta} of (\ref{Eq:16}) (\textit{i.e.}, the fermionic Green's functions). 
\par
For the sake of an easier presentation, it is convenient to proceed with a simplified derivation of effective locality.

\subsection{Effective locality at eikonal and quenching approximations}

Without loss of generality, things can be illustrated on the basis of a 4-point fermionic Green's function. Then, two propagators $G_F(x_1,y_1|A)$ and $G_F(x_2,y_2|A)$ are to be represented with the help of a \emph{Fradkin representation} such as (mixed case),
\begin{eqnarray}\label{Gfrad}
{\langle p|{G}_{F}[A] |y \rangle} &=&  e^{-i p \cdot y} \  i
\int_{0}^{\infty}\mathrm{d}s \, e^{-is m^2} \ e^{- \frac{1}{2} \mathrm{Tr}\,{\ln{\left( 2h \right)}} }\nonumber\\ \nonumber && \times \int{\mathrm{d}[u]} {\left\{ m - i \gamma \cdot \left[p - g A(y-u(s)) \right] \right\}} \ e^{\frac{i}{4} \int_{0}^{s}{\mathrm{d}s' \, [u'(s')]^{2} } } \ e^{i p \cdot u(s)}\nonumber\\  && \times  \left( e^{g \int_{0}^{s}{\mathrm{d}s' \sigma \cdot \mathbf{F}(y-u(s'))}} \ e^{-ig \int_{0}^{s}{\mathrm{d}s' \, u'(s') \cdot \mathbf{A}(y-u(s'))}} \right)_{+}\,,
\end{eqnarray}with,
\begin{equation} h(s_1,s_2)=s_1\Theta(s_2-s_1)+s_2\Theta(s_1-s_2)\ ,\ \ \ h^{-1}(s_1,s_2)=\frac{\partial}{\partial s_{1}} \frac{\partial}{\partial s_{2}} \delta(s_1-s_2).\end{equation}
In (\ref{Gfrad}), the 4-vector $u(s)$ is the Fradkin variable, while in the last line the $+$-subscript indicates an $s'$-Schwinger-proper-time ordering of the expression between parenthesis. Clearly, Fradkin's representations cannot be thought of as being of a very simple usage. But they are exact. Recently, they have even been used to re-derive some standard results of Quantum Mechanics in a certainly involved but successful way \cite{RefM}.
\par
Upon substitution into (\ref{Eq:18}), a representation like (\ref{Gfrad}) produces a cumbersome structure of an exponential of an exponential. This is why it is necessary to bring (\ref{Gfrad}) down by means of functional differentiations with respect to the Grassmannian sources $\bar{\eta},\eta$ to deal with $2n$-point fermionic Green's functions. Accordingly, in contrast to the field strength formulation dual to the pure Yang Mills case \cite{RefF}, the property of effective locality is referred to the behavior of fermionic Green's functions rather than that of the generating functional itself. Of course, this is not a restriction of generality because fermionic Green's functions exhaust the whole fermionic content of the theory.
\par
In (\ref{Gfrad}) though, the linear and quadratic $A^a_\mu$-field dependences are contained within a time-ordered exponential which prevents the linkage (\ref{link}) to operate in a simple way. This can be circumvented at the expense of introducing two extra field variables so as remove the $A^a_\mu$-field variables from the ordered exponential. For example, this can be achieved by writing,
\begin{equation}\label{out}
\left(e^{ig\,p^\mu\!\int_{-\infty}^{+\infty} {\rm{d}}s\,A^a_\mu(y-sp)\,T^a}\right)_+=\mathcal{N}\int {\rm{d}}[\alpha]\int{\rm{d}}[\Omega]\, e^{-i\!\int_{-\infty}^{+\infty} {\rm{d}}s\,\,\Omega^a(s)[\alpha^a(s)-gp^\mu A^a_\mu(y-sp)]}\left(e^{i\int_{-\infty}^{+\infty}{\rm{d}}s\,\alpha^a(s)T^a}\right)_+
\end{equation}where $\mathcal{N}$ is a normalisation constant. Equation (\ref{out}) defines an \emph{eikonal} approximation of (\ref{Gfrad}) which is used here to offer a simple derivation of the property of interest. Likewise, in this subsection, the approximation of quenching is applied, and the Fradkin representation of the functional $L(A)$ can be ignored. 
\par
Now, not until the full integrations on the two extra field variables $\alpha^a$ and $\Omega^a$ are brought to completion, can one be assured to deal with the proper Fradkin representation of $G_F[A]$. At this level of approximations though and in the strong coupling limit, $g\gg 1$, it is a fortunate circumstance that these extra dependences can be integrated out in an exact way thanks to the \emph{Random Matrix} calculus \cite{{QCD6},{RefI}}.
\par
Omitting for short the details of integrations on Schwinger proper times, Fradkin's variables and the four extra fields $\alpha^a_i$ and $\Omega^a_i$, $i=1,2$, one obtains a result of the following form
\begin{eqnarray}\label{ELEQ}
\prod_{i=1}^{2}\int\mathrm{d}s_i \int\mathrm{d}u_i(s_i)\int\mathrm{d}\alpha_i(s_i)\int\mathrm{d}\Omega_i(s_i)\,\left(\ddots\right) \int{\mathrm{d}[\chi] \, e^{ \frac{i}{4} \int{ \chi^{2} }} } \, \left. e^{\mathfrak{D}_{A}^{(0)}} \, e^{{+ \frac{i}{2} \int{ A ^a_\mu\, K^{\mu\nu}_{ab}\, A^b_\nu} }} \, e^{i\int{Q^a_\mu A^\mu_a } }\right|_{\ A \rightarrow 0 }\,,
\end{eqnarray}
where $K^{\mu\nu}_{ab}$ and $Q^a_\mu$ represent the quadratic and linear $A^a_\mu$-field dependences, respectively,
\begin{equation}\label{QK}
K_{\mu\nu}^{ab}=gf^{abc}\chi_{\mu\nu}^c+\left({{D}_{\mathrm{F}}^{(0)}}^{-1}\right)_{\mu \nu}^{a b}\,, \ \ \ Q^a_\mu =-\partial^\nu\chi^a_{\mu\nu}+g[R^a_{1,\mu}+R^a_{2,\mu}]\,, \ \  \ f^{abc}\chi^c_{\mu\nu}=
(f\cdot \chi)^{ab}_{\mu\nu}\,,\end{equation}and where the $R^a_{i,\mu}$ stand for the part of (\ref{out}) linear in the potential field $A^a_\mu$,
 \begin{equation}\label{currents}
 R^a_{i,\mu}(z) = p_{i,\mu}\, \int\mathrm{d}s_i\,\Omega^a_{i}(s_i)\,\delta^4(z - y_{i} + s_i\, p_{i})\,, \ \ \ \ \ i=1,2\,. 
 \end{equation} Note that in (\ref{currents}) the eikonal approximation has substituted $s_ip_i$ for the 
 original Fradkin field variable $u_i(s_i)$ ({\textit{i.e.}} a straight line approximation, connecting the points $z$ and $y_i$).
The linkage operation followed by the prescription of cancelling the sources $j^a_\mu$ is now trivial and yields,
\begin{eqnarray}
& & \left. e^{-\frac{i}{2} \int{\frac{\delta}{\delta A} \cdot {D}_{\mathrm{F}}^{(0)} \cdot  \frac{\delta}{\delta A} }} \cdot e^{+ \frac{i}{2} \int{A \cdot {K} \cdot A} + i \int{A \cdot {Q} }}  \right|_{A \rightarrow 0} \\ \nonumber &=& e^{-\frac{1}{2} \Tr{\ln{\left( 1- {D}_{\mathrm{F}}^{(0)} \cdot {K} \right)}}} \cdot e^{\frac{i}{2} \int{{Q} \cdot \left[ {D}_{\mathrm{F}}^{(0)} \cdot \left( 1 - {K} \cdot {D}_{\mathrm{F}}^{(0)}\right)^{\!-\!1} \right]\cdot {Q}}}.
\end{eqnarray}

\noindent On the right hand side, the kernel of the quadratic term in ${Q}_{\mu}^{a}$ is
\begin{eqnarray}\label{magics}
{D}_{\mathrm{F}}^{(0)} \cdot \left( 1 - {K} \cdot {D}_{\mathrm{F}}^{(0)}\right)^{\!-\!1} &=& {D}_{\mathrm{F}}^{(0)} \cdot \left( 1 - \left[ g f \cdot \chi + {{D}_{\mathrm{F}}^{(0)}}^{\!-\!1} \right] \cdot {D}_{\mathrm{F}}^{(0)}\right)^{\!-\!1} \\ \nonumber &=& - \left( g f \cdot \chi \right)^{\!-\!1},
\end{eqnarray}so that eventually,
\begin{eqnarray}\label{EL}
& & \left. e^{-\frac{i}{2} \int{\frac{\delta}{\delta A} \cdot {D}_{\mathrm{F}}^{(0)} \cdot  \frac{\delta}{\delta A} }} \cdot e^{+ \frac{i}{2} \int{A \cdot {K} \cdot A} + i \int{A \cdot {Q} }}  \right|_{A \rightarrow 0} \\ \nonumber &=& e^{-\frac{1}{2} \Tr{\ln{\bigl[-g{D}_{\mathrm{F}}^{(0)}\bigr]}}} \cdot \frac{1}{\sqrt{\det(f\cdot\chi)}}\cdot e^{-\frac{i}{2} \int\mathrm{d}^4z\ {{Q}(z) \cdot (gf\cdot\chi(z))^{-1}\cdot {Q}(z)}}\,,
\end{eqnarray}where the first term is a constant (possibly infinite) which can be absorbed into a redefinition of the overall normalization constant $\mathcal{N}$. 
\par
The manifestation of effective locality is in the last term of (\ref{EL}). While the $\left({\mathbf{D}_{\mathrm{F}}^{(0)}}^{-1}\right)_{\mu \nu}^{a b}$-piece of $K^{\mu\nu}_{ab}$ in (\ref{QK}) is \emph{non-local} it disappears from the final result so as to leave the \emph{local} structure $[gf\cdot\chi]^{-1}$ as the mediator of 
interactions between quarks,
\begin{equation}\label{local}
\langle x | (gf\cdot\chi)^{-1} | y \rangle = (gf\cdot\chi)^{-1}(x) \, \delta^{(4)}(x-y)\,.
\end{equation} This offers a simple way to look at the effective locality phenomenon whose detailed derivation can be seen to rely in an essential way on the non-abelian character of the gauge group. As already transparent at the level of equations (\ref{Eq:9}), in effect, and contrary to expectations \cite{RefO} this phenomenon cannot show up in the abelian case of QED because the $\mathcal{L}'[A]$ interaction of (\ref{glue}) doesn't exist in that case. Going back to (\ref{ELEQ}), one finds a result whose final form reads as 
\begin{eqnarray}\label{ELEQ'}
\mathcal{N}\,\prod_{i=1}^{2}\int\mathrm{d}s_i \int\mathrm{d}u_i(s_i)\int\mathrm{d}\alpha_i(s_i)\int\mathrm{d}\Omega_i(s_i)\,\left(\ddots\right) \int{\mathrm{d}[\chi] \, e^{ \frac{i}{4} \int{ \chi^{2} }} } \nonumber\\ \cdot\frac{1}{\sqrt{\det(f\cdot\chi)}}\cdot e^{-\frac{i}{2} \int\mathrm{d}^4z\ {{Q}(z) \cdot (gf\cdot\chi(z))^{-1}\cdot {Q}(z)}}\,,\end{eqnarray}
and it is the integration on the $\chi^a_{\mu\nu}$-fields which, thanks to another remarkable consequence of effective locality, lends itself to a(n analytically continued \cite{QCD6}) Random Matrix calculation \cite{{QCD6},{RefI}}.
\par
Through more cumbersome expressions, the above structure generalizes easily to $2n$-point fermionic Green's functions, see \cite{RefI} (Appendix D). As mentioned already in the Introduction, in the pure (euclidean) YM case, a form of effective locality was observed in an attempt at dualization of the YM theory \cite{RefF}. 


\subsection{Effective locality in general }
All simplifications leading to Eqs.(\ref{magics})-(\ref{ELEQ'}) can be relaxed without affecting effective locality \cite{QCD-II}. 
Quenching corresponds to the replacement $ \det {\backslash\!\!\!\!{D}}(A)=e^{L(A)}\longmapsto 1$ while the eikonal approximation entails the consistent neglect
of the spin contribution in the time-ordered exponential of (\ref{Gfrad}), the term $\exp[{g \int_{0}^{s}{\mathrm{d}s' \sigma \cdot \mathbf{F}(y-u(s'))}}]$, as well as, for the Fradkin's field variables, the replacement of $u_i(s_i)$ by its \emph{straight-line} approximation $s_ip_i$, with $i=1,2,\dots,n$, in the case of a $2n$-point fermionic Green's function. For the sake of a simpler illustration, though, the case of a 4-point fermionic Green's function will be considered again. One has, 

\begin{eqnarray}\label{Eq:19}
& & {M}(x_{1}, y_{1}; x_{2}, y_{2})\nonumber  \\  & & = \left. \frac{\delta}{\delta \bar{\eta}(y_{1})} \cdot \frac{\delta}{\delta \eta(x_{1})} \cdot \frac{\delta}{\delta \bar{\eta}(y_{2})} \cdot \frac{\delta}{\delta \eta(x_{2})} \cdot \mathcal{Z}_{}\left\{ j, \bar{\eta}, \eta \right\} \right|_{\eta=\bar{\eta}=0; j=0}\nonumber  \\  & & = \, \mathcal{N} \, \int{d[\chi] \, e^{\frac{i}{4}\int{\chi^{2}}} \, e^{\mathfrak{D}_{A}^{(\zeta)}} \,} e^{ + \frac{i}{2}\int{\chi\cdot \mathbf{F}} +\frac{i}{2}\int{A \cdot \left(\mathbf{D}_{F}^{(\zeta)}\right)^{-1} \cdot A }} \, \left. {G}_{\mathrm{F}}(x_{1}, y_{1}|gA) \, {G}_{\mathrm{F}}(x_{2}, y_{2}|gA) \, e^{{L}[A]} \right|_{A=0}\nonumber  \\  & & \quad \quad \quad - \{ {1} \leftrightarrow {2} \},
\end{eqnarray}where the anti-symmatrization indicated in the last line is redundant in the non-approximated situation but necessary in case of approximations \cite{Hugo's}.

	In order to display the gauge variance of all such QCD correlation functions, it is appropriate to recall that the fermionic loop functional ${L}[A]$ is explicitly invariant under the full $SU(3)_c$ gauge group of QCD.  Then, it is convenient to combine the Gaussian $A$-dependence of every entering of ${G}_{\mathrm{F}}[A]$ into a net expression,
\begin{eqnarray}\label{Eq:20}
\exp{\left[ \frac{i}{2}\int{\mathrm{d}^{4}z \, A^{\mu}_{a}(z) \, {{K}}_{\mu \nu}^{ab}(z) \, A^{\nu}_{b}(z)} + i\int{\mathrm{d}^{4}z \, {{Q}}^{\mu}_{a}(z) \, A_{\mu}^{a}(z)} \right]},
\end{eqnarray}

\noindent where ${{K}}$ and ${{Q}}$ are local functions of the Fradkin variables, $u^{\mu}_i(s_i)$, and of auxiliary fields $\Omega_1^{a}(s_{1})$, $\Omega_2^{a}(s_{2})$. Because of the spin-terms, extra fields, $\Phi^a_{1,\mu \nu}(s_{1})$, $\Phi^a_{2,\mu \nu}(s_{2})$, are necessary to extract the $A_{\mu}^{a}(y - u(s'))$ potentials from the full ordered exponential of (\ref{Gfrad}) as can be recognized in a representation like ($i=1,2$)
\begin{eqnarray}\label{full}
& & \left( e^{ -ig \int_{0}^{s_i}{ds_i' \, {u'}_i^{\mu}(s_i') \, A_{\mu}^{a}(y_i-u_i(s_i')) \, \lambda^{a}} + g \int_{0}^{s_i}{ds_i' \sigma^{\mu \nu} \, {F}_{\mu \nu}^{a}(y_i-u_i(s_i')) \, \lambda^{a}}} \right)_{+} \\ \nonumber &=& \mathcal{N}_{\Omega} \, \mathcal{N}_{\Phi} \, \int{d[\alpha_i] \, \int{d[{\Xi_i}] \, \int{d[\Omega_i] \, \int{d[\Phi_i] \, \left( e^{ i \int_{0}^{s_i}{ds_i' \, \left[ \alpha_i^{a}(s_i') - i \sigma^{\mu \nu} \, {\Xi_i}_{\mu \nu}^{a}(s_i') \right] \, \lambda^{a}}} \right)_{+} }}}} \\ \nonumber & & \quad \times\, e^{-i \int_0^{s_i}{ds_i' \, \Omega_i^{a}(s_i') \, \alpha_i^{a}(s_i')}  - i \int_0^{s_i}{ds_i' \, {\Phi_i}^{a\mu \nu}(s_i') \,  {\Xi_i}^{a}_{\mu \nu}(s_i') } } \\ \nonumber & & \quad \times\, e^{- i g \int_0^{s_i}{ds'_i \, u'^{\mu}(s_i') \, \Omega_i^{a}(s_i') \, A^{a}_{\mu}(y_i-u_i(s_i')) } + i g \int_0^{s_i}{ds_i' \, {\Phi_i}^{a}_{\mu \nu}(s_i') \, {F}^{\mu \nu}_{a}(y_i-u_i(s_i'))}  }.
\end{eqnarray}
 For the 4-point function, the product of ${G}_{\mathrm{F}, \mathrm{1}}[A]$ and ${G}_{\mathrm{F}, \mathrm{2}}[A]$ will contribute to the kernel ${{K}}_{\mu \nu}^{ab}(z)$ the following the terms
\begin{eqnarray}\label{spins}
{{K}}_{\mu \nu}^{ab}(z) = & & 2 g^{2} \int_{0}^{s_{1}}{\mathrm{d}s_1' \ \delta^{(4)}(z - y_{1} + u_1(s_1')) f^{abc}\, \Phi_{\mu\nu,\mathrm{1}}^{c}(s_1')}\nonumber \\  & & + 2 g^{2} \int_{0}^{s_{2}}{\mathrm{d}s_2' \ \delta^{(4)}(z - y_{2} + {u_2}(s_2')) f^{abc}\, \Phi_{\mu\nu,\mathrm{2}}^{c}(s_2')}\,,
\end{eqnarray}

\noindent and to the `current' ${{Q}}_{\mu \nu}^{a}(z)$, the terms, 
\begin{eqnarray}\label{Eq:22}
{{Q}}_{\mu}^{a}(z) = & & - 2g\, \partial^{\nu} \Phi_{\mathrm{I}, \nu \mu}^{a}(z) - g \int_{0}^{s_{1}}{\mathrm{d}s_1' \, \delta^{(4)}(z - y_{1} +u_1(s_1')) \, u'_{1,\mu}(s_1')\, \Omega_{\mathrm{1}}^{a}(s_1')} \nonumber\\  & & - 2g\, \partial^{\nu} \Phi_{\mathrm{2},\nu \mu}^{a}(z) - g \int_{0}^{s_{2}}{\mathrm{d}s_2' \, \delta^{(4)}(z - y_{2} + {u_2}(s_2')) \, u'_{2,\mu}(s_2')\, \Omega_{\mathrm{2}}^{a}(s_2')}\,,
\end{eqnarray}where, out of (\ref{full}), one has
\begin{eqnarray}\label{Eq:23}
\Phi_{i,\mu \nu}^{a}(z) \equiv \int_{0}^{s_i}{\mathrm{d}s_i' \, \delta^{(4)}(z - y_i +u_i(s_i')) \, \Phi_{i,\mu \nu}^{a}(s_i')}\,.
\end{eqnarray}Dropping the first terms in the two right hand sides of (\ref{Eq:22}) and proceeding to the eikonal replacement of $u_i(s_i)$ by $s_ip_i$, one recovers the eikonal `currents' of (\ref{currents}). In the case of higher $2n$-point fermionic Green's functions additional terms will of course complete the ${{Q^a_\mu}}$ and ${{K}}_{\mu \nu}^{ab}$ functionals in exactly the same way.
 \par\medskip\smallskip
 Then, by relaxing the quenching approximation, the required operation reads 
\begin{eqnarray}\label{fullinkage}
\exp{\left[- \frac{i}{2} \int{\frac{\delta}{\delta A} \cdot {D}_{\mathrm{F}}^{(0)} \cdot \frac{\delta}{\delta A}}\right]} \cdot \exp{\left[ \frac{i}{2}\int{A \cdot \bar{{K}} \cdot A} + i \int{\bar{{Q}} \cdot  A} \right]} \cdot \exp{\left({L}[A] \right)},
\end{eqnarray}followed by the prescription of cancelling the potentials $A^a_\mu$, and where, 
\begin{eqnarray}\label{fullK}
\langle z | \bar{{K}}_{\mu \nu}^{ab} |z' \rangle = \left[ {{K}}_{\mu \nu}^{ab}(z) + g f^{abc} \chi_{\mu \nu}^{c}(z) \right] \, \delta^{(4)}(z - z') + \langle z | \left. \left({D}_{F}^{(0)} \right)^{-1} \right|_{\mu \nu}^{ab} |z' \rangle\,,
\end{eqnarray}
\begin{eqnarray}\label{fullQ}
\bar{{Q}}_{\mu}^{a}(z) =\partial^{\nu} \chi_{\nu \mu}^{a}(z)+ {{Q}}_{\mu}^{a}(z)\,.
\end{eqnarray}
In $\bar{{K}}$ all terms but the inverse of the gluon propagator are local. Equation (\ref{fullinkage}) requires the linkage operator to act upon the product of two functionals of $A$. This is accounted for by a functional identity,
\begin{eqnarray}\label{doublelink}
e^{\mathfrak{D}_{A}} \, \mathcal{F}_{1}[A] \mathcal{F}_{2}[A] = \left( e^{\mathfrak{D}_{A}} \, \mathcal{F}_{1}[A] \right) \,  e^{\overleftrightarrow{\mathfrak{D}}} \, \left( e^{\mathfrak{D}_{A'}} \, \mathcal{F}_{2}[A'] \right)
\end{eqnarray}
where the 'cross-linkage' operator $\exp\{\overleftrightarrow{\mathfrak{D}}\}$ is defined in the following way,
\begin{equation}\label{crosslinkage}
\overleftrightarrow{\mathfrak{D}} = -i \int{\overleftarrow{\frac{\delta}{\delta A}}  {D}_{\mathrm{F}}^{(0)} \overrightarrow{\frac{\delta}{\delta A'}}}\,,
\end{equation}and where in (\ref{doublelink}) the limit $A^a_\mu=A'^a_\mu=0$ is understood after execution of the linkage operations.
With the identifications,
\begin{eqnarray}\label{F1F2}
\mathcal{F}_{1}[A] =  \exp{\left[ \frac{i}{2}\int{A \cdot \bar{{K}} \cdot A} + i \int{\bar{{Q}} \cdot  A} \right]}, \quad\quad   \mathcal{F}_{2}[A] = \exp{\left( \mathbf{L}[A] \right)}
\end{eqnarray}
the evaluation of $e^{\mathfrak{D}_{A}} \, \mathcal{F}_{1}[A]$ reads
\begin{eqnarray}\label{LinkF1}
e^{\mathfrak{D}_{A}} \, \mathcal{F}_{1}[A] &=& \exp{\left[ \frac{i}{2} \int{ \bar{{Q}} \cdot {D}_{F}^{(0)} \cdot (1 - \bar{{K}} \cdot {D}_{F}^{(0)})^{-1} \cdot \bar{{Q}}} -\frac{1}{2} \Tr{\ln{\left(1-{D}_{F}^{(0)} \cdot \bar{{K}}\right)}} \right]} \\ \nonumber & & \quad \cdot \exp{\left[ \frac{i}{2} \int{ A \cdot \bar{{K}} \cdot \left( 1 - {D}_{F}^{(0)} \cdot \bar{{K}} \right)^{-1}\! \cdot A} + i \int{\bar{{Q}} \cdot \left( 1- \bar{{K}} \cdot {D}_{F}^{(0)} \right)^{-1}\! \cdot A} \right]}\,,
\end{eqnarray}
where the same cancellation of non-local pieces as observed in (\ref{magics}) is taking place again,
\begin{equation}\label{magics'}
{D}_{F}^{(0)} \cdot \left( 1 - \bar{{K}} \cdot {D}_{F}^{(0)} \right)^{-1} = {D}_{F}^{(0)}  \cdot \left[ 1 -  (\widehat{{K}}  + \left({D}_{F}^{(0)}\right) ^{-1}) \cdot {D}_{F}^{(0)} \right]^{-1} = - {\widehat{{K}}}^{-1},
\end{equation}
and where the `gluon bundle' term of $(gf\cdot\chi)^{-1}$ of (\ref{local}) is now completed by the spinorial contributions of (\ref{spins})
\begin{eqnarray}\label{Eq:32}
\widehat{{K}}_{\mu \nu}^{ab} = {{K}}_{\mu \nu}^{ab} + g f^{abc} \chi_{\mu \nu}^{c}\,.
\end{eqnarray}

Now, in the limit $A \rightarrow 0$ the crossed-linkage operation of (\ref{doublelink}) first yields,
\begin{eqnarray}\label{Eq:33}
e^{\mathfrak{D}_{A}} \, \mathcal{F}_{1}[A] \mathcal{F}_{2}[A]_{|_{A=0}} = & & \exp{\left[ -\frac{i}{2} \int{\bar{{Q}} \cdot \widehat{{K}}^{-1} \cdot \bar{{Q}}} - \frac{1}{2} \Tr\ln{\widehat{{K}}} +  \frac{1}{2} \Tr\ln{\left(-{D}_{\mathrm{F}}^{(0)}\right)} \right]} \\  & & \cdot \exp{\left[+ \frac{i}{2} \int{\frac{\delta}{\delta A'} \cdot {D}_{\mathrm{F}}^{(0)} \cdot \frac{\delta}{\delta A'}}\right]} \cdot \exp{\left[ \frac{i}{2} \int{\frac{\delta}{\delta A'} \cdot \widehat{{K}}^{-1} \cdot \frac{\delta}{\delta A'}} - \int{\bar{{Q}} \cdot \widehat{{K}}^{-1} \cdot \frac{\delta}{\delta A'} }\right]}\nonumber \\ \nonumber & & \cdot \left( e^{\mathfrak{D}_{A'}} \, \mathcal{F}_{2}[A'] \right),
\end{eqnarray}followed by the prescription $A'=0$. The first exponential term in the second line of (\ref{Eq:33}) is exactly $\exp{\{- \mathfrak{D}_{A'}\}}$ and removes the $\exp{\{\mathfrak{D}_{A'}\}}$ of the operation $\left(\exp{\{\mathfrak{D}_{A'}\}} \cdot \mathcal{F}_{2}[A']\right)$.  With the exception of an irrelevant $\exp{\left[\frac{1}{2}\Tr\ln{\left(-{D}_{\mathrm{F}}^{(0)}\right)}\right]}$ factor, to be absorbed into an overall normalization, what remains to all orders of coupling and for every such process is therefore the following generic structure, 
\begin{eqnarray}\label{surprising}
e^{\mathfrak{D}_{A}} \, \mathcal{F}_{1}[A] \mathcal{F}_{2}[A]_{|_{A=0}} = & & \mathcal{N} \, \exp{\left[ -\frac{i}{2} \int{\bar{{Q}} \cdot \widehat{{K}}^{-1} \cdot \bar{{Q}}} - \frac{1}{2} \Tr\ln{\widehat{{K}}} \right]} \\ \nonumber & & \cdot  \left.\exp{\left[ \frac{i}{2} \int{\frac{\delta}{\delta A} \cdot \widehat{{K}}^{-1} \cdot \frac{\delta}{\delta A}} - \int{\bar{{Q}} \cdot \widehat{{K}}^{-1} \cdot \frac{\delta}{\delta A} }\right]} \cdot \exp{\left( {L}[A] \right)} \right|_{A\rightarrow 0}
\end{eqnarray}

\noindent in which the prime in $A'$ has been suppressed.  
\par\medskip\noindent
In (\ref{surprising}) it is clear that nothing refers to ${D}_{\mathrm{F}}^{(0)}$: Gauge invariance is achieved as a matter of gauge independence. Assuming, as in the previous quenched situation, that integrations over the extra fields $\Xi$ and $\Phi$ can be carried through, this conclusion is non-trivial. It should be obvious at this stage that exactly the same conclusion is reached whatever the field function in use, ${D}_{\mathrm{F}}^{(0)}$, ${D}_{\mathrm{F}}^{(\zeta)}$, ${D}_{\mathrm{F}}^{(\mathbf{n})}$, etc..  \par

Again, a striking aspect of (\ref{surprising}) is its locality. Still, (\ref{surprising}) doesn't provide the bases of a genuinely dual formulation of full QCD. For example, the most famous duality correspondence of $g\rightarrow 1/g$ cannot account for the various $g$-scaling behaviours that are exhibited by (\ref{surprising}) whereas it does so in the pure Yang Mills case \cite{RefF}.
 
 \subsection{Effective locality even more generally }
 The derivations of effective locality that have been displayed so far rely on Fradkin's representations and exhibit both locality and gauge-fixing independence. A natural question to ask is whether effective locality turns out to be representation dependent. According to the preceding developments it is sufficient to examine this point in the quenched case since the functional $L[A]$ is itself local and gauge-invariant. Likewise, for this very purpose, it is possible to restrict our consideration to the case of a fermionic 2-point function without restriction of generality. The expression to be considered thus is
 \begin{equation}
 \mathcal{N}  \int d[\chi] \, e^{\frac{i}{4}\int \chi^2} \, \left. e^{\mathfrak{D}_A^{(o)}} \, e^{ + \frac{i}{2}\int{\chi\cdot {F}} +\frac{i}{2}\int{A \cdot \left({D}_{F}^{(0)}\right)^{-1}\!\! \cdot\, A }} \, {G}_{\mathrm{F}}(x_{}, y_{}|A) \right|_{A\rightarrow 0}\end{equation}with no particular representation assumed for $G_F[A]$. This is 
 \begin{equation}
 \mathcal{N}  \int d[\chi] \, e^{\frac{i}{4}\int \chi^2} \, \left. e^{\mathfrak{D}_A^{(0)}} \, e^{ + \frac{i}{2}\int{A\cdot K\cdot A}\, +\,i\int A^\mu_a\, \partial^\nu\chi^a_{\nu\mu} }\ G_{\mathrm{F}}(x, y|A) \right|_{A\rightarrow 0}\end{equation}at,
 \begin{equation}
 K=g f^{abc} \chi_{\mu \nu}^{c}+\left. \left({D}_{F}^{(0)} \right)^{-1} \right|_{\mu \nu}^{ab}\,, \end{equation}so that, with the new identifications,
 \begin{eqnarray}\label{F'1F'2}
\mathcal{F}_{1}[A] =  \exp\left[ \frac{i}{2}\int A \cdot K \cdot A + i \int A^\mu_a\, \partial^\nu\chi^a_{\nu\mu} \right], \quad\quad   \mathcal{F}_{2}[A] =G_{\mathrm{F}}(x,y|A)\end{eqnarray}the double operation of (\ref{doublelink}) yields, 
\begin{eqnarray}\label{surprising'}
e^{\mathfrak{D}_{A}} \, \mathcal{F}_{1}[A] \mathcal{F}_{2}[A]_{|_{A=0}} = & & \mathcal{N}e^{\frac{1}{2}\Tr\ln{\left(-{D}_{\mathrm{F}}^{(0)}\right)}} \, \exp{\left[ -\frac{i}{2} \int{{{\nabla\chi}} \cdot {{(gf\cdot\chi)}}^{-1} \cdot {{\nabla\chi}}} - \frac{1}{2} \Tr\ln{{{(gf\cdot\chi)}}} \right]} \\ \nonumber & & \cdot  \left.\exp{\left[ \frac{i}{2} \int{\frac{\delta}{\delta A} \cdot {{(gf\cdot\chi)}}^{-1} \cdot \frac{\delta}{\delta A}} - \int{{{\nabla\chi}} \cdot {{(gf\cdot\chi)}}^{-1} \cdot \frac{\delta}{\delta A} }\right]} G_{\mathrm{F}}(x,y|A) \right|_{A\rightarrow 0}\,,
\end{eqnarray}where the shorthand $(\nabla\chi)^a_\mu=\partial^\nu\chi^a_{\nu\mu}$ was used.
\par\medskip
 Again, any gauge-fixing dependence has disappeared from the result. Effective locality is manifest in the first line of (\ref{surprising'}) and, if the new functional $\mathcal{F}_2=G_{\mathrm{F}}(x,y|A)$ is not gauge-invariant itself, the full second line of (\ref{surprising'}) is. It is straightforward to check that the same applies to a product of $n$ functionals, $\prod_{i=1}^n G_{\mathrm{F}}(x_i,y_i|A)$, that is to the case of a generic $2n$-point fermionic Green's function. 
 \par
 While Fradkin's representations are useful to actual calculations and in derivations of other remarkable properties \cite{{QCD6},{RefI}} the gist of effective locality (see below) doesn't depend on them.

  \section{\label{SEC:4}Conclusion: The meaning of effective locality}
  Established on the basis of formal, yet standard functional manipulations effective locality emerges as a non-perturbative property of QCD fermionic Green's functions which is not bound to any approximation scheme \textit{artefact}.
  \par
   Since its discovery some years ago, effective locality has been able to display a number of promising {\textit{tree-level}} consequences \cite{QCD1,QCD-II,QCD5, QCD6, QCD5', RefI} while its meaning has remained enigmatic. At variance with the pure Yang Mills euclidean case \cite{RefF}, in effect, this property cannot be thought of in terms of an expression that would be dual to the usual form of QCD.
   \par
   Deriving effective locality along the lines of the current paper, an interesting disconnectedness is observed between the gauge field functions (\textit{i.e.}, propagators) and the corresponding gauge fixing terms. The clue to this situation comes from the fact that any gauge field propagator disappears from the final result so as to leave gauge-fixing independent expressions. This suggests the following interpretation and/or conjecture concerning the meaning of effective locality.
   \par
   Effective locality is the very mode in which non-abelian gauge invariance is realized in the non-perturbative regime of QCD.
   \par\noindent
   Contrarily to the well controlled perturbative regime where, up to substantial complications, the non-abelian gauge invariance is as indirect as it is in the abelian case of QED, gauge independence of amplitudes enjoys a much more direct and thus simpler mode of realization in the non-perturbative regime of QCD. That is, effective locality indicates how the non-perturbative realization of local gauge-invariance in QCD differs from its perturbative QED-like realization. This is testified by the widely recognized inherently perturbative character of the BRS(T) symmetry \cite{Becchi}. A related, suggestive indication of this is the issue of \textit{Gribov copies} which has recently been declared even worse than it was initially thought to be, and eventually intractable \cite{Thess}. 
  In this perspective it appears reasonable to think that the Gribov's copies issue indicates that one is leaving the perturbative regime, and that its \emph{intractability} is the indication that the non-perturbative regime of QCD cannot be attained in this way.   Now, such an obstruction doesn't plague the effective locality non-perturbative framework since no gauge-fixing is ever introduced in this approach.
   \par
   One may note that while asymptotic electron/positron states correspond to plain elementary physical particles in QED,  the only QCD asymptotic states that are outright \emph{physical} are not quarks/antiquarks states, but those of the hadronic spectrum. Beyond the matter of its formal derivation, this remark is in order to render the surprising effective locality statement of (\ref{surprising}) (or (\ref{surprising'})) somewhat more `moral'/acceptable : This astonishingly direct mode of non-abelian gauge invariance realization has to do with the non-perturbative regime of QCD, that is, with the only QCD regime giving rise to genuine  physical states. In this respect, QCD would again appear simpler and more self-consistent \cite{wilczek} a theory than QED.
   \par
   Other deep aspects of effective locality show up as exact representations of the fermionic field functionals are implemented in the course of concrete Green's functions calculations, and will be the matter of a forthcoming publication \cite{htr2}.
 \par

%
%
%

\end{document}